\documentclass[aps,pra,twocolumn,superscriptaddress]{revtex4}
\usepackage{color}
\usepackage{epsfig}
\usepackage{graphicx}
\usepackage{amssymb}
\usepackage{amsmath}
\usepackage{amsthm}
\usepackage{hyperref}

\newcommand{\nn}{\nonumber \\}
\newcommand{\bra}[1]{\left\langle #1\right|}
\newcommand{\ket}[1]{\left| #1\right\rangle}
\newcommand{\ketbra}[3][]{\left|#2\right\rangle_{#1}\!\left\langle#3\right|}
\newcommand{\eeqref}[1]{Eq.~(\ref{#1})}
\newcommand{\tr}[0]{{\rm Tr}}
\newcommand{\Es}{E_{\rm s}}
\newcommand{\Ep}{E_{\rm u}}
\newcommand{\Ed}{E_{\rm d}}
\newcommand{\nomod}{K}
\newcommand{\nomodf}{u}
\newcommand{\noout}{M}

\begin{document}

\title{Preservation of loss in linear-optical processing}

\author{Dominic W. Berry}
\affiliation{Institute for Quantum Computing, University of Waterloo, Waterloo, Ontario N2L 3G1, Canada}
\author{A. I. Lvovsky}
\affiliation{Institute for Quantum Information Science, University of Calgary, Alberta T2N 1N4, Canada}

\begin{abstract}
We propose a measure of quantum efficiency of a multimode state of light that quantifies the amount of optical loss
this state has experienced, and prove that this efficiency cannot increase in any linear-optical processing with
destructive conditional measurements. Any loss that has affected a state can neither be removed nor redistributed so
as to further increase the efficiency in higher-efficiency modes at the expense of lower-efficiency modes. This
result eliminates the possibility of catalytically improving photon sources.
\end{abstract}


\maketitle

\section{Introduction}
A leading approach to quantum information processing is via linear-optical quantum computing (LOQC),
first proposed in 2001 by Knill, Laflamme and Milburn \cite{KLM}. Major progress has been made both on the
theoretical and experimental fronts towards implementation of LOQC. Modifications have been proposed that greatly
reduce the overhead costs \cite{KokRMP}, a quantum error correction protocol has been introduced
\cite{Dawson06,Varnava}, and experimental implementation of primary gates has been demonstrated \cite{KLMexp}.

In spite of this progress, practical LOQC is still out of reach. Many of the difficulties arise because the
single-photon sources required for LOQC, as well as computational circuits themselves, suffer from losses. Although a
certain degree of tolerance to losses does exist in some LOQC schemes \cite{Varnava}, the efficiency of existing
single-photon sources \cite{NJPissue} as well as the quality of individual circuit elements and waveguides are far
below the required minima.

Under these circumstances it appears beneficial to develop a procedure that would reverse the effect of losses,
perhaps at a cost of introducing extra resources. It would be useful, for example, to employ the outputs of $N$
imperfect single-photon sources to obtain $\nomod<N$ single-photon sources of improved quantum efficiency. Accomplishing this task
would be straightforward if nonlinear-optical interactions with single photons were readily available: for example, one could employ non-demolition photon number measurements to select only those modes that contain photons.
However, achieving such interactions is extremely technically challenging
\cite{KimblePhotonGate}, whereas linear-optical (LO) processing is easily achieved in the laboratory.

It is therefore important to investigate whether elimination of losses is possible under
LO processing. Under this processing we understand arbitrary interferometric transformations and
conditioning on results of arbitrary \emph{destructive} measurements on some of the optical modes involved. The
efforts to construct such a scheme began in 2004, mostly ending with various no-go results
\cite{Berry04a,Berry04b,Berry06,Berry07}. The most general result to date was obtained in Ref.~\cite{BL}.
In that work, we quantified the \emph{efficiency} of a quantum optical state by the amount of loss that state might have experienced.
We then proved that the efficiency in any single-mode optical state obtained through LO processing cannot exceed the quantum efficiency of the best available single-mode input \cite{BL}.

However, those previous results had limited application to multimode states.
First, as we show below, extending the definition of the efficiency of a quantum state to the multimode case is not straightforward, particularly when the loss has been ``mixed" among the modes by interferometric transformations.
Second, our earlier results do not provide any information on how the efficiencies can be distributed among the output modes, aside from the general upper bound mentioned above.
For example, they leave open the possibility of a ``catalytic'' scheme, in which some high-efficiency single photons are used to obtain additional high-efficiency single photons.

In the present work we generalize our study of the dynamics of optical losses to the multimode case. We introduce the notion of quantum
efficiency of a (possibly entangled) multimode state which quantifies the amount of loss this state may have
experienced. We show that this efficiency cannot increase under LO processing. That is, any loss that has occurred at
the input can neither be removed nor redistributed so as to improve the efficiency in some of the modes at the
expense of lower-efficiency modes. This means that there is a majorization relation between the efficiencies at the input and the output. The LO processing can act to average the efficiencies, but not to concentrate them. This rules out, in particular, any possibility of catalytic efficiency improvement.

\section{Single-mode measures of efficiency}
Before describing multimode measures of efficiency, we discuss the properties and relationships for single-mode measures of efficiency that have been previously proposed.
Usually efficiency is used to describe a process for producing a state.
However, it is also convenient to regard efficiency as a measure on the state itself, regardless of the process used to produce it \cite{Berry04a,Berry04b,Berry06,Berry07,BL}.
Specifically, Ref.~\cite{BL} uses the efficiency to quantify \emph{the maximum amount of loss an optical mode carrying the given state might have previously experienced}:
\begin{equation}\label{mostgen}
E(\hat\rho) := \inf \left\{ p |\ \exists \hat \rho_0\ge 0 ~:~ {\cal E}_{p} (\hat \rho_0)=\hat \rho \right\},
\end{equation}
where ${\cal E}_{p}$ is a loss channel with transmissivity $p$. That is, one considers all hypothetical methods of producing the given state $\hat\rho$  via loss from some valid initial quantum state $\hat\rho_0$.

We emphasize that the loss is just a way of mathematically quantifying the efficiency of the state.
It is not necessary that the state were created by such a process.
The efficiency is a measure on the state, and should not be regarded as an intrinsic feature of the mode.

Let us study a few examples. Ideally, single photon sources would produce a single photon state $\ket 1$ on demand.
In practice, such sources may with some probability fail to produce a photon, and there is no way to detect this failure without destructive measurement. Therefore the state produced by a generic single photon source may be approximated as
\begin{equation}
\label{eq:mix}
\hat\rho = p \ket{1}\bra{1} + (1-p) \ket{0}\bra{0}.
\end{equation}
Here the quantity $p$ is commonly referred to as the efficiency of the single photon source. In the context of our definition, state \eqref{eq:mix} can be obtained from the single-photon state by transmitting a (perfect) single photon through a loss channel with transmissivity $p$, and hence its efficiency equals $p$. In this way, the efficiency of state \eqref{eq:mix} according to our new definition is consistent with the traditional definition of the efficiency of a single-photon source.

Coherent states have efficiency exactly equal to zero, regardless of their amplitude.
This is because coherent states remain coherent states under loss.
A coherent state of amplitude $\alpha$ can be obtained from one of amplitude $\alpha/\sqrt{p}$ under a loss channel of transmissivity $p$. Although one cannot take $p=0$ (because complete loss always results in the vacuum state), possible values of $p$ form an open set with zero infimum.

On the other hand, any pure state \emph{other} than a coherent state (or the vacuum state) must have efficiency 1.
This is because a state under loss is a mixture of the original state, and the state with different numbers of photons lost.
That is, a pure state $\ket{\chi}$ becomes a mixture of $\ket{\chi}$, $\hat a\ket{\chi}$, $\hat a^2\ket{\chi}$, and so forth.
The only way in which the state after loss can remain pure is if $\ket{\chi}\propto\hat a\ket{\chi}$.
The only states for which this is true are eigenstates of the annihilation operator; i.e.\ coherent states.

Determining the efficiency of a known single-mode state is a straightforward computational task.
The loss channel ${\cal E}_{p}$ corresponds to a linear transformation known as the generalized Bernoulli transformation. 
Provided the state $\hat\rho$ can be obtained via loss channel ${\cal E}_{p}$ from some initial operator, we can define the inverse map ${\cal E}_{p}^{-1}$, which can be calculated as in Ref.\ \cite{Herzog}.
Therefore, we need to find the infimum of the values of $p$ such that the inverse Bernoulli mapping ${\cal E}_{p}^{-1}(\hat\rho)$ exists and yields a valid quantum state, i.e.\ can be represented by a positive semidefinite density matrix.

A further interesting feature of a state's efficiency is that it equals zero if and only if the state is classical, i.e.\ it can be written as a statistical mixture of coherent states, or, equivalently, its Glauber-Sudarshan $P$ function has the properties of a probability density. As discussed above, any coherent state has efficiency zero, and hence so does any statistical mixture  of coherent states. To prove the converse, let us suppose there exists a nonclassical state $\hat\rho$ such that  $E(\hat\rho)=0$. Let $\Phi_{\hat\rho}(\eta)$ denote the Fourier transform of this state's $P$ function $P(\alpha)$ over the phase space. According to Bochner's theorem \cite{Bochner}, because $P(\alpha)$ is not a probability density, there exist two sets of $n$ complex numbers $\eta_k$ and $z_k$, such that
\begin{equation}\label{bochnereq}
\sum\limits_{i,j=1}^n \Phi_{\hat\rho}(\eta_i-\eta_j)z_iz_j^*<0.
\end{equation}

Because $E(\hat\rho)=0$, for any $p>0$ there exists state $\hat\rho_0$ such that $\hat \rho$ is obtained from $\hat\rho_0$ by means of attenuation by factor $p$. Because attenuation corresponds to ``shrinkage" of the $P$ function in the phase space \cite{Leonhardt}, we have  $\Phi_{\hat\rho_0}(\eta)=\Phi_{\hat\rho}(\eta/\sqrt p)$ and hence
\begin{equation}\label{bochnereq0}
\sum\limits_{i,j=1}^n \Phi_{\hat\rho_0}(\eta'_i-\eta'_j)z_iz_j^*<0,
\end{equation}
where $\eta'_k=\eta_k\sqrt p$. By choosing $p$ close to zero, the set of arguments of  function $\Phi_{\hat\rho_0}$ in the above equation can be upper bounded by an arbitrarily small value $A$.

Now recall that the Husimi $Q$ function of any quantum state must be non-negative. This means that the Fourier transform $\Psi_{\hat\rho_0}(\eta)$ of the $Q$ function of state $\hat\rho_0$ must obey
\begin{equation}\label{bochnerpsi}
\sum\limits_{i,j=1}^n \Psi_{\hat\rho_0}(\eta'_i-\eta'_j)z_iz_j^*\ge 0.
\end{equation}
But the $Q$ function is obtained from the $P$ function by convolving the latter with a Gaussian, $e^{-|\alpha|^2}/\pi$ \cite{Leonhardt}. This means that the Fourier transforms of these functions are connected by multiplication,
\begin{equation}\label{}
\Psi_{\hat\rho}(\eta)=\Phi_{\hat\rho}(\eta)e^{-|\eta|^2}.
\end{equation}
By choosing $p$ close to zero, one can make the factor $e^{-|\eta|^2}$ arbitrarily close to 1 within radius $A$. Accordingly, the left-hand sides of Eqs.~\eqref{bochnereq0} and \eqref{bochnerpsi} are equal in the limit $p\to 0$. We arrive at a contradiction, which means that any nonclassical state  $\hat \rho$ must have a finite efficiency $E(\hat\rho)>0$.

\section{Multimode measures of efficiency}
\label{sec:mul}
Let us now generalize the notion of efficiency to an optical state carried by multiple modes. A direct generalization can be obtained by assuming that each mode has propagated through its own loss channel, and taking the sum of the transmissivities:
\begin{equation}\label{ED}
\Ed(\hat\rho,\nomod) := \inf \left\{ \sum_{\ell=1}^\nomod p^\downarrow_{\ell}\ |\ \exists \hat \rho_0\ge 0 ~:~ {\cal E}_{\vec p} (\hat \rho_0)=\hat \rho \right\}.
\end{equation}
The notation $p^\downarrow_\ell$ indicates the elements of the vector $\vec p$ sorted in non-increasing order. The value of $\nomod$ can be less than the number of modes constituting state $\hat\rho$. In this way, the efficiency is defined not only for the entire state, but also for a subset of $\nomod$ modes with the lowest losses. This extension facilitates comparison of efficiencies of states with different number of modes.

A drawback of this definition is that it does not adequately take account of loss that has been mixed between modes. For example, consider two polarization modes carrying a single-photon qubit in the state $\ket{\psi}=\ket{1_H}\ket {0_V}$. The efficiency of the state in the horizontally polarized mode is 1, and that in the vertically polarized mode 0, so $\Ed(\ketbra\psi\psi,2)=1$. On the other hand, writing the same state in terms of diagonal polarization modes, we find $\ket{\psi'}=(\ket{1_{+45^\circ}}\ket{0_{-45^\circ}}+\ket{0_{+45^\circ}}\ket{1_{-45^\circ}})/\sqrt 2$. This state cannot be obtained by independent loss in the two modes, and would have a different efficiency, $\Ed(\ketbra{\psi'}{\psi'},2)=2$, even though its utility for quantum information processing is exactly the same as that of $\ket\psi$.

An alternative approach to quantifying the efficiency is to treat each mode separately, and calculate the sum of single-mode efficiencies for $\nomod$ highest-efficiency modes:
\begin{equation}\label{Ei}
\Es(\hat\rho,\nomod):= \sum_{\ell=1}^\nomod E(\tr_{\forall k\ne \ell}\hat\rho)^\downarrow.
\end{equation}
This definition is also problematic. First, similarly to the d-efficiency \cite{FootNoteNom}, it depends on the choice of the mode basis. For the example above, $\Es(\ketbra\psi\psi,1)=1$, but $\Es(\ketbra{\psi'}{\psi'},1)=1/2$. Second, it may underestimate the efficiency in many cases. For example, the s-efficiency of the state  $\ket\phi=\sqrt{1-p}\ket{00}+\sqrt{p}\ket{11}$ equals $\Es(\ketbra\phi\phi,1)=p$, and can be very small. On the other hand, conditioning on detection of a photon in one of the modes of $\ket\phi$ results in a perfect single photon in the other mode, as is the case with producing heralded single photons via parametric down-conversion. State $\ket\phi$ is thus much more useful than, for example, single-mode state $\hat\sigma=(1-p)\ketbra{0}{0}+p\ketbra{1}{1}$, which has the same s-efficiency but cannot be processed to produce a high-quality single photon.

We aim to provide a definition of efficiency that would be invariant with respect to transformation of modes and adequately reflect the state's value for quantum information purposes.
To this end we modify the definition $\Ed$ by including an optimization over interferometers.
That is, we consider simultaneous loss channels on each of the modes ${\cal E}_{\vec p}$, followed by an arbitrary interferometer $W$, as shown in Fig.\ \ref{fig:int1}(a).
The efficiency is then the sum of the $\nomod$ largest values of $p_\ell$:
\begin{equation}
\label{eq:kef} \Ep(\hat\rho,\nomod):=  \inf \left\{ \!\sum_{\ell=1}^\nomod p^\downarrow_{\ell} \Big| \exists
\hat \rho_0\ge 0, W :W{\cal E}_{\vec p} (\hat \rho_0)=\hat \rho \right\}.
\end{equation}

An important property of the u-efficiency \eqref{eq:kef} is its invariance with respect to interferometric transformation of modes. Indeed, if state $\hat\rho'$ can be obtained from state $\hat\rho$ by applying interferometric transformation $U$, so that $\hat\rho'=U\hat\rho$, and we have $W{\cal E}_{\vec p} (\hat \rho_0)=\hat \rho$ in the context of \eeqref{eq:kef}, we also have $UW{\cal E}_{\vec p} (\hat \rho_0)=\hat \rho'$. But transformation $UW$ can be treated as a single interferometer, which means that $\Ep(\hat\rho',\nomod)\le\Ep(\hat\rho,\nomod)$. But because interferometric transformations are reversible, we also have $\Ep(\hat\rho,\nomod)\le\Ep(\hat\rho',\nomod)$ and hence $\Ep(\hat\rho',\nomod)=\Ep(\hat\rho,\nomod)$.

Similar to the case for the efficiency $E$, the \nomodf-efficiency can be calculated via inverting the channel.
In finite dimension, the channel given by the loss followed by the unitary operation $W$ may be represented by a matrix, which may be inverted to find $\hat\rho_0$.
The efficiency can then be found by a minimization over $\vec p$ and $W$ such that $\hat\rho_0$ is a valid quantum state.
For the other two efficiencies, the calculation is simpler.
For the d-efficiency, one only needs to minimize over $\vec p$, and for the s-efficiency one can just determine the single-mode efficiencies for the reduced density matrices in the individual modes.

Let us evaluate the multimode efficiency of the example states studied above. State $\ket\psi$ is a tensor product and has $\Es(\ketbra\psi\psi,2)=\Es(\ketbra\psi\psi,1)=1$. As we show below, the d-, s-, and u-efficiencies coincide for tensor product states, so we also have $\Ep(\ketbra\psi\psi,2)=\Ep(\ketbra\psi\psi,1)=1$. Since the u-efficiency is invariant under interferometric transformations, state $\ket{\psi'}$ has the same u-efficiency. Analyzing each of the modes of state $\ket{\psi'}$ on its own, we find them to carry the state  $(\ket{1}\bra{1} + \ket{0}\bra{0})$/2, so  $\Es(\ketbra{\psi'}{\psi'},1)=1/2$ and $\Es(\ketbra{\psi'}{\psi'},2)=1$. For state $\ket\phi$, both the u- and d-efficiencies equal 2. This is because, even if subjected to an interferometric transformation, it is a pure state that is not coherent, and hence cannot be obtained by attenuating another state.

\section{Proof that the \nomodf-efficiency cannot increase under LO processing}
\begin{figure}[b]
\center{\includegraphics[width=\columnwidth]{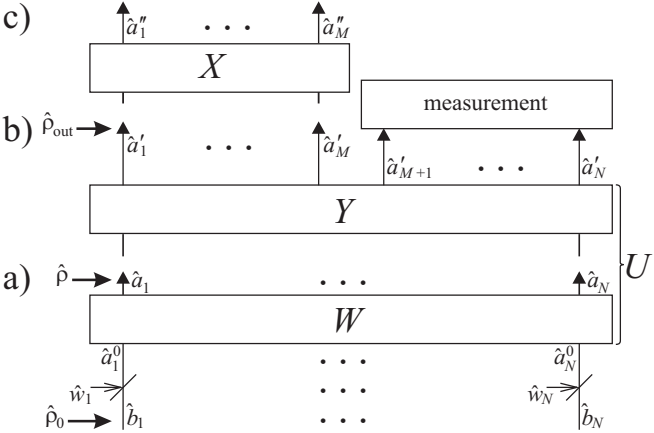}} \caption{\label{fig:int1} A general setup for LO processing.
(a) To determine the efficiency of the input state $\hat\rho$, we find an initial state $\hat\rho_0$, such that $\hat\rho$ may be obtained by attenuation and interferometer $W$ according to \eeqref{eq:kef}.
(b) LO processing of the input state.
The modes pass through a general interferometer, and all but $\noout$ of the output modes are detected via a measurement.
The state $\hat\rho_{\rm out}$ of the remaining $\noout$ modes can be conditioned on a particular measurement result.
(c) The upper limit on the efficiency of the output state is established by choosing an interferometer, $X$, through which this state can be transmitted such that the resulting state, carried by modes $\hat a''_m$, can be obtained by multimode attenuation of another state.}
\end{figure}

In this section we show that it is impossible to increase the \nomodf-efficiency using LO processing.
A general LO scheme is shown in Fig.\ \ref{fig:int1}(b).
The input state $\hat\rho$, carried by $N$ optical modes with annihilation operators $\hat a_1,\ldots,\hat a_N$, is passed through a general interferometer which performs a unitary operation $Y$ on these mode operators.
We retain $\noout$ of the output modes $\hat a'_i$, and the remaining $N-\noout$ modes are subjected to a generalized destructive quantum measurement.
We consider postselection on a particular result of this measurement, and determine the \nomodf-efficiency of the state $\hat\rho_{\rm out}$ carried by the remaining output modes.
Our goal is to prove that
\begin{equation}\label{noincr}
\Ep(\hat\rho_{\rm out},\nomod)\le \Ep(\hat\rho,\nomod)
\end{equation}
for any $\nomod\le \noout$.

In accordance with definition \eqref{eq:kef}, we model the state $\hat\rho$ as being obtained from some initial state $\hat\rho_0$ by combining each of its modes, $\hat b_j$, with vacuum $\hat w_j$ on a beam splitter with transmissivity $p_j$ [Fig.\ \ref{fig:int1}(a)]:
\begin{equation}
\hat a^0_j = \sqrt{p_j} \hat b_j + \sqrt{1-p_j} \hat w_j,
\end{equation}
followed by interferometer $W$.
We assume that the settings are chosen such that, for some $\epsilon>0$,
\begin{equation}\label{Ep}
\sum_{\ell=1}^\nomod p^\downarrow_\ell \le \Ep(\hat\rho,\nomod)+\epsilon.
\end{equation}
The introduction of $\epsilon$ takes account of the possibility that there does not exist a setting which achieves the infimum.

Because interferometers $W$ and $Y$ are adjacent to each other, we can without loss of generality treat them as a single interferometer, corresponding to unitary transformation $U=YW$. The action of this interferometer can be written as
\begin{equation}\label{IntAct}
\hat a'_i = \sum_{j=1}^N U_{ij} \hat a^0_j =\sum_{j=1}^N U_{ij}\sqrt{p_j} \hat b_j + \sum_{j=1}^N U_{ij}\sqrt{1-p_j}
\hat w_j.
\end{equation}
We see that each vacuum mode contributes to each of the output modes, including those that are subjected to conditional measurements. These measurements may ``compromise" the vacuum contributions to the output state \cite{FootNoteComp}, so the output efficiency cannot be calculated directly from the matrix elements $U_{ij}$. We address this issue by performing an RQ decomposition on the matrix
$U_{ij}\sqrt{1-p_j}$ such that
\begin{equation}\label{Rmat}
U_{ij}\sqrt{1-p_j} = \sum_{\ell=1}^N R_{i\ell} Q_{\ell j},
\end{equation}
where $Q$ is unitary and $R$ is an upper triangular matrix, so $R_{i\ell}=0$ for $\ell<i$. Then we get
\begin{equation}
\label{eq:out} \hat a'_i = \sum\limits_{\ell=1}^N U_{i\ell}\sqrt{p_\ell} \hat b_\ell + \sum\limits_{\ell=1}^N
R_{i\ell} \hat v_\ell,
\end{equation}
where
\begin{equation}
\hat v_\ell := \sum_{j=1}^N Q_{\ell j}\hat w_j
\end{equation}
are obtained by transforming modes $\hat w_j$ in a fictitious interferometer $Q$. Because all the $\hat w_j$
correspond to vacuum states, so do the $\hat v_\ell$. The subset $\{\hat v_1,\ldots,\hat v_M\}$ of these modes does not contribute to the set of output modes $\{\hat a'_{M+1},\ldots, \hat a'_N\}$ that is subjected to measurement, and thus directly leads to the loss of efficiency in the output state.

Without loss of generality, we append another interferometer, $X$, acting on the $\noout$ output modes.
Because the \nomodf-efficiency is independent of linear interferometers, this interferometer does not affect the \nomodf-efficiency at the output.
To determine the interferometer to use, we perform a singular value decomposition on the upper left $\noout\times \noout$ block of $R$ such that
\begin{equation}\label{XRpQp}
R = X^\dagger R' Q',
\end{equation}
where the upper left $\noout\times \noout$ block of $R'$ is diagonal, and unitaries $X$ and $Q'$ are equal to the identity outside the upper left $\noout\times \noout$ block.
We choose the unitary matrix $X$ for the final interferometer to be that given by this decomposition.

Denoting the annihilation operators for the modes after the interferometer $X$ by $\hat a''_k$, we have, for $k\le\noout$,
\begin{align}
\label{eq:out2} & \hat a''_k = \sum_{i=1}^\noout X_{ki} \hat a'_i \nn
&=\sum_{i=1}^\noout X_{ki} \left( \sum_{\ell=1}^N U_{i\ell}\sqrt{p_\ell} \hat b_\ell
+ \sum_{\ell=1}^N R_{i\ell} \hat v_\ell \right) \nn
&=\sum_{i=1}^\noout \sum_{\ell=1}^N X_{ki}U_{i\ell}\sqrt{p_\ell} \hat b_\ell
+\sum_{i=1}^\noout \sum_{\ell=\noout+1}^N X_{ki} R_{i\ell} \hat v_\ell \nn
& \quad + \sum_{i=1}^\noout \sum_{\ell=1}^\noout X_{ki} \sum_{k',n=1}^\noout [X^\dagger]_{ik'} R'_{k'n} Q'_{n\ell} \hat v_\ell \nn
&= \sum_{i=1}^K\sum\limits_{\ell=1}^N X_{ki} U_{i\ell}\sqrt{p_\ell} \hat b_\ell
+ \sum_{i=1}^\noout \sum_{\ell=\noout+1}^N X_{ki} R_{il} \hat v_{\ell}
+ R'_{kk} \hat v''_k,
\end{align}
where
\begin{equation}
\hat v''_k := \sum_{\ell=1}^N Q'_{k\ell} \hat v_\ell.
\end{equation}

As the set $\{\hat v''_k\}$ may be regarded as being obtained from initial vacuum modes $\{\hat w_k\}$ via a unitary transformation, they represent an orthonormal set of bosonic modes in the vacuum state. Furthermore, those $\hat v''_k$ that contribute to $\hat a''_k$ do not contain any contribution from the ``compromised" vacuum modes. Indeed, they only contain contributions from $\hat v_\ell$ for $\ell\le M$, whereas the operators for the measured modes only contain contributions from $\hat v_\ell$ for $\ell>M$. As a result, these vacuum contributions are equivalent to loss.

\begin{figure}[b]
\center{\includegraphics[width=\columnwidth]{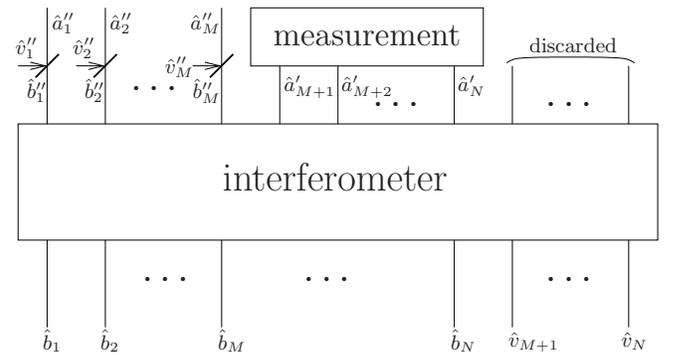}} \caption{\label{fig:int2} A rearrangement of the interferometer.
The vacuum modes $\{\hat v''_1,\ldots,\hat v''_\noout,\hat v_{\noout+1}\ldots\hat v_N\}$ can be obtained via an interferometer (not shown) from the original vacuum modes $\hat w_\ell$.
The modes $\hat b_\ell$ and $\{\hat v_{\noout+1},\ldots,\hat v_N\}$ are combined in the interferometer to produce $\{\hat b''_1,\ldots,\hat b''_\noout\}$, as well as $\{\hat a'_{\noout},\ldots,\hat a'_N\}$, which are measured, and some modes which are discarded.
The modes $\{\hat b''_1,\ldots,\hat b''_\noout\}$ are then combined with vacuum modes $\{\hat v''_1,\ldots,\hat v''_\noout\}$ to generate the output state.}
\end{figure}

To make this result explicit, we write the annihilation operator in the form $\hat a''_k = \hat B''_k + \hat
V''_k$, where
\begin{equation}
\hat B''_k = \sum_{i=1}^\noout\sum\limits_{\ell=1}^N X_{ki} U_{i\ell}\sqrt{p_\ell} \hat b_\ell + \sum_{i=1}^\noout
\sum_{\ell=\noout+1}^N X_{ki} R_{il} \hat v_{\ell} ,
\end{equation}
and
\begin{equation}
\label{eq:vdef} \hat V''_k = R'_{kk} \hat v''_k.
\end{equation}
We then find that
\begin{align}
[\hat V''_k , (\hat V''_{k'})^\dagger ] &= \delta_{kk'}|R'_{kk}|^2, \\ [\hat B''_k , (\hat B''_{k'})^\dagger ] &=
\delta_{kk'}(1-|R'_{kk}|^2).
\end{align}
The first line follows immediately from Eq.\ \eqref{eq:vdef}. The second line is obtained because $\hat B''_k =
\hat a''_k - \hat V''_k$ and $[\hat a''_k , (\hat a''_{k'})^\dagger ] = \delta_{kk'}$.
Defining
\begin{equation}
p''_k := 1-|R'_{kk}|^2, \qquad \hat b''_k := \hat B''_k/\sqrt{p''_k},
\end{equation}
we have
\begin{equation}
\hat a''_k = \sqrt{p''_k} \hat b''_k + \sqrt{1-p''_k}\hat v''_k.
\end{equation}
Therefore, the output state may be obtained by an interferometer that produces the modes with annihilation
operators $\hat b''_k$, then combining with vacua on beam splitters with transmissivities $p''_k$, as shown in Fig.\ \ref{fig:int2}.


Without loss of generality, we can assume $X$ and $Q'$ to have been chosen such that the numbers $p''_k$ are in non-increasing
order.
The \nomodf-efficiency at the output is therefore upper bounded by
\begin{equation}\label{Epout}
\Ep(\hat\rho_{\rm out},\nomod) \le \sum_{i=1}^\nomod p''_k.
\end{equation}
To determine the sum \eqref{Epout}, we can define the unitaries
\begin{equation}
U' := XU, \qquad Q'' := Q' Q.
\end{equation}
It follows from \eeqref{XRpQp} that $R' = X R (Q')^\dagger$. Therefore, according to \eeqref{Rmat},
\begin{equation}
R'_{k\ell} = \sum_{m=1}^N U'_{km} \sqrt{1-p_m} (Q''_{\ell m})^*.
\end{equation}
Then we obtain
\begin{align}
&\sum_{k=1}^\nomod p''_k \le \nomod-\sum_{k=1}^\nomod\sum_{\ell=1}^\nomod |R'_{k\ell}|^2 \nn
&= \nomod-\sum_{k=1}^N\sum_{\ell=1}^\nomod\sum_{m,j=1}^N U'_{km}\sqrt{1-p_m} (Q''_{\ell m})^* \nn &\quad \times (U'_{kj})^*\sqrt{1-p_j} Q''_{\ell j} \nn
&= \nomod-\sum_{\ell=1}^\nomod\sum_{m,j=1}^N \delta_{m j}\sqrt{1-p_m} (Q''_{\ell m})^*\sqrt{1-p_j} Q''_{\ell j} \nn
&= \nomod-\sum_{\ell=1}^\nomod\sum_{j=1}^N (1-p_j) |Q''_{\ell j}|^2 \nn &= \sum_{\ell=1}^\nomod\sum_{j=1}^N p_j |Q''_{\ell j}|^2 \le \sum_{\ell=1}^\nomod p^\downarrow_\ell.
\label{longproof}
\end{align}
The last inequality in \eeqref{longproof} is obtained because $Q''_{ij}$ is unitary, so $|Q''_{ij}|^2$ is a doubly stochastic matrix, and thus vector $p_l$ majorizes vector \cite{NielsenChuang}
\begin{equation}
q_\ell := \sum_{j=1}^N p_j |Q''_{ij}|^2.
\end{equation}

Now, according to Eqs.~\eqref{Ep}, \eqref{Epout}, \eqref{longproof} and because we can choose $\epsilon$ to be arbitrarily close to zero, we obtain
\begin{equation}
\label{EKleEK}
\Ep(\hat\rho_{\rm out},\nomod) \le \Ep(\hat\rho,\nomod).
\end{equation}
This is the main result of this work: the universal measure of quantum efficiency of a multimode state, the
\nomodf-efficiency, cannot increase under LO processing.

\section{Comparison of efficiency measures}
We now use the above result to prove some additional properties of the different measures of multimode efficiency defined in Sec.~III. First, we show that these efficiencies are related according to
\begin{equation}\label{EffIneq}
\Es(\hat\rho,\nomod)\le \Ep(\hat\rho,\nomod)\le \Ed(\hat\rho,\nomod).
\end{equation}
To examine the s-efficiency, we can again assume that state $\hat\rho$ is obtained via a set of beam splitters with transmissivities
$p_j$ and an interferometer $W$ as in Fig.\ \ref{fig:int1}(a), such that the sum of the $\nomod$ largest values of
$p_j$ is no more than $\Ep(\hat\rho,K)+\epsilon$.
Then the operators for the state $\hat\rho$ are given by
\begin{equation}
\hat a_j = \hat B_j + \hat V_j,
\end{equation}
with
\begin{equation}
\hat B_j :=\sum_{\ell=1}^N W_{j\ell}\sqrt{p_\ell}\hat b_\ell, \qquad \hat V_j :=\sum_{\ell=1}^N
W_{j\ell}\sqrt{1-p_\ell}\hat w_\ell,
\end{equation}
corresponding to operators carrying signal and vacuum fields, respectively.

To determine the s-efficiency, we determine the efficiency for each mode individually.
When determining the efficiency for mode $\hat a_j$, we can regard modes $\hat a_k$ for $k\ne j$ as being discarded.
The vacuum operators $\hat V_k$ for $k\ne j$ are not orthogonal to $\hat V_j$; however, since those modes are discarded, the addition of vacuum $\hat V_j$ is equivalent to loss.
Therefore, the efficiency of the state in mode $\hat a_j$ is no greater than
\begin{equation}
p'_j:=[\hat B_j,\hat B_j^\dagger]=\sum_{\ell=1}^N |W_{j\ell}|^2 p_\ell.
\end{equation}
The sum of the $K$ largest values of $p'_j$ upper bounds the s-efficiency; that is,
\begin{equation}
\Es(\hat\rho,\nomod)\le \sum_{j=1}^\nomod {p'}^\downarrow_j.
\end{equation}
Because $W$ is unitary, $|W_{j\ell}|^2$ is a doubly stochastic matrix, and the vector of values $\vec p$
majorizes $\vec p'$. That means that
\begin{equation}
\sum_{j=1}^\nomod {p'}^\downarrow_j \le \sum_{j=1}^\nomod p_j^\downarrow \le \Ep(\hat\rho,\nomod)+\epsilon.
\end{equation}
Because this holds for all $\epsilon>0$, we have $\Es(\hat\rho,\nomod)\le \Ep(\hat\rho,\nomod)$.

The second inequality in \eeqref{EffIneq} is because the definition of $\Ep(\hat\rho,\nomod)$ in \eeqref{eq:kef} looks for the minimum in a larger set of states than that of $\Ed(\hat\rho,\nomod)$ in \eeqref{ED}.

For tensor product states, the s- and d-efficiencies are the same.
To see this, use the definition \eqref{Ei} on the tensor product state
\begin{equation}
\hat \rho = \bigotimes_{j=1}^N \hat\rho_j.
\end{equation}
One obtains
\begin{equation}
\Es(\hat\rho,\nomod)= \sum_{\ell=1}^\nomod E(\hat\rho_{\ell})^\downarrow.
\end{equation}
Therefore, there exists a set of states $\hat\rho^0_j$ and transmissivities $p_j$, such that the sum of the $\nomod$ largest values of $p_j$ is no more than $\Es(\hat\rho,\nomod)+\epsilon$, and the final states $\hat\rho_j$ may be obtained via loss channels with transmissivities $p_j$ from initial states $\hat\rho^0_j$.
This would also provide a scheme for producing $\hat\rho$ for the definition of $\Ed(\hat\rho,\nomod)$, and therefore
\begin{equation}
\Ed(\hat\rho,\nomod) \le \sum_{j=1}^\nomod p_j^\downarrow \le \Es(\hat\rho,\nomod)+\epsilon.
\end{equation}
Because this is true for all $\epsilon>0$, we obtain $\Ed(\hat\rho,\nomod) \le \Es(\hat\rho,\nomod)$.
Combining this with \eeqref{EffIneq}, we find that $\Ed(\hat\rho,\nomod) = \Es(\hat\rho,\nomod)$ for tensor product states, and all inequalities in \eqref{EffIneq} saturate.

This result leads us to an important conclusion. Suppose we start with $N$ separable states (for example, imperfect single photons as in \eeqref{eq:mix}) with efficiencies $p_\ell$, which we subject to LO processing, resulting in a set of modes in which the efficiencies, when analyzed separately, are given by $p'_\ell$.
Using the result that LO processing cannot increase the u-efficiency, and \eeqref{EffIneq}, we have for any integer $K$,
\begin{equation}
\label{eq:cat}
\sum_{\ell=1}^\nomod p'_\ell \le \sum_{\ell=1}^\nomod p^\downarrow_\ell .
\end{equation}
In other words, the LO processing can act to average the efficiencies, but not to concentrate them.
One consequence is the exclusion of any possibility for ``catalytic'' efficiency improvement, in which some
highly efficient sources are used to increase the efficiency in other optical modes, without themselves suffering
from loss.

These results do not rule out increases in the individual efficiencies; for example, if the largest efficiency is decreased, it is possible for the second largest efficiency to be increased.

\section{Summary}
We have introduced a number of measures that enable us to quantify the efficiency in multimode systems.
The \nomodf-efficiency is a powerful general measure that takes account of how loss may have been mixed between the different modes.
It is unchanged under linear interferometers, and cannot increase under more general LO processing with destructive measurements.
We have used this result to show that catalytic improvement of photon sources is not possible with LO processing.
If one starts with independent optical sources (which produce a tensor product of states), then the efficiencies in the individual output modes are weakly majorized by the efficiencies in the input.
This means that it is not possible to concentrate the efficiencies, such that the sum of the highest $\nomod$ output efficiencies is greater than the sum of the highest $\nomod$ input efficiencies.

It is clearly possible to increase the \nomodf-efficiency if one uses \emph{non}linear optical elements.
For example, a standard method of producing single photons is via parametric downconversion (a nonlinear process), and postselection on detection of a photon in one of the output modes.
The initial beam is coherent, with efficiency zero, but the final output (ideally) has unit efficiency.

\acknowledgments
This work has been supported by NSERC, AIF, CIFAR and Quantum{\it Works}. We thank B.\ C.\ Sanders and H.\ M.\ Wiseman for stimulating
discussions.


\begin{thebibliography}{10}
\bibitem{FootNoteNom} We will call $E_d$ the d-efficiency, $E_s$ the s-efficiency, and so on.
\bibitem{FootNoteComp} Consider, for example, a single photon in mode $\hat a$ overlapped with vacuum $\hat v$ on a beam splitter. One of the beam splitter output modes, $(\hat a-\hat v)/\sqrt2$, is subjected to a photon number measurement, and the state of the second output mode is analyzed conditioned on the result of the measurement. The annihilation operator of the latter mode is given by $(\hat a+\hat v)/\sqrt2$, so one may argue that its efficiency cannot exceed $1/2$ due to the vacuum contribution. However, if we condition on detecting the vacuum state in the first mode, the state of the second mode is a single photon, i.e.\ it has efficiency 1, in spite of the vacuum contribution. In this context, we say that the vacuum contribution of the second mode is ``compromised" by the conditional measurement in the first one.
\bibitem{footnote3} If $p''_m$ equals $0$ or $1$, modes $\hat b''_m$ and $\hat v''_m$, respectively, are not defined. In these cases we have, respectively, $\hat a''_m=\hat v''_m$ or $\hat a''_m=\hat b''_m$.
\end{thebibliography}
\end{document}